# Interfacial charge transfer and interaction in the MXene/TiO$_2$ heterostructures

Lihua Xu,[1,2,†], Tao Wu,[2,†] Paul R. C. Kent,[3,4] and De-en Jiang[1,2,*]

[1]Department of Chemical and Environmental Engineering, University of California, Riverside, California, 92521, USA

[2]Department of Chemistry, University of California, Riverside, California, 92521, USA

[3]Computational Sciences and Engineering Division, Oak Ridge National Laboratory, Oak Ridge, Tennessee 37831, USA

[4]Center for Nanophase Materials Sciences, Oak Ridge National Laboratory, Oak Ridge, Tennessee 37831, USA

*E-mail: djiang@ucr.edu

**Abstract:**

Hybrid materials of MXenes (2D carbides and nitrides) and transition-metal oxides (TMOs) have shown great promise in electrical energy storage and 2D heterostructures have been proposed as the next-generation electrode materials to expand the limits of current technology. Here we use first principles density functional theory to investigate the interfacial structure, energetics, and electronic properties of the heterostructures of MXenes (Ti$_{n+1}$C$_n$T$_2$; T=terminal groups) and anatase TiO$_2$. We find that the greatest work-function differences are between OH-terminated-MXene (1.6 eV) and anatase TiO$_2$(101) (6.4 eV), resulting in the largest interfacial electron transfer (~0.9 e/nm$^2$ across the interface) from MXene to the TiO$_2$ layer. This interface also has the strongest adhesion and further strengthened by hydrogen bond formation. For O-, F-, or mixed O-/F- terminated Ti$_{n+1}$C$_n$ MXenes, electron transfer is minimal and interfacial adhesion is weak for their heterostructures with TiO$_2$. The strong dependence of the interfacial properties of the MXene/TiO$_2$ heterostructures on the surface chemistry of the MXenes will be useful to tune the heterostructures for electric-energy-storage applications.

## I. INTRODUCTION

Great efforts have been devoted to enhancing the performance of electrical energy storage devices in the past decade. However, it is still challenging to achieve high energy density and power density simultaneously [1–4]. Novel electrode materials hold the key to improved performance. Due to their excellent electrical conductivity and high volumetric capacitance, MXenes (2D carbides and nitrides) were explored in applications for electrical energy storage and electrocatalysis [5–12]. Especially, many researchers have demonstrated great potentials of using MXenes for capacitive energy storage [13–19].

To further enhance the performance, MXene nanosheets such as $Ti_3C_2T_x$ (T denotes terminal groups such as -O, -OH, and -F) were hybridized with transition metal oxides (TMOs), and the resultant binder-free flexible films exhibited excellent Li-ion storage capability [20,21]. For example, Rakhi et al. reported that the nanocrystalline ε-$MnO_2$ coated with MXene nanosheets (ε-$MnO_2$/$Ti_2CT_x$ and ε-$MnO_2$/$Ti_3C_2T_x$) showed superior specific capacitance than the pure MXene-based symmetric supercapacitors [22]. In another report, Ahmed et al. successfully fabricated $TiO_2$ nanocrystals on the surface of $Ti_2CT_x$ MXene sheets for Li-ion battery applications [23]. Heterostructures of MXene and graphene have also attracted great interest [24–30].

Despite the recent experimental demonstrations of using MXene/TMO composite/hybrid materials as the electrodes for energy storage [31–33], first principles understanding of the interfacial structure, energetics, and electronic properties is still missing. $Ti_3C_2T_x$ is the prototypical MXene and metallic, but its thinner cousin $Ti_2CT_x$ offers higher specific capacitance (due to smaller formula weight) despite its semiconducting nature [34,35]. Thus, it will be interesting to examine the contrast between $Ti_2CT_x$ and $Ti_3C_2T_x$ in forming interfaces with TMOs, with potential applications for electric energy storage.

Herein we investigate the interfacial properties of the $Ti_2CT_x$/$TiO_2$ and $Ti_3C_2T_x$/$TiO_2$ heterostructures using the first principles density-functional theory (DFT), as an initial step toward understanding the MXene/TMO interfaces and heterostructures. We chose $Ti_3C_2T_x$ because it is the most studied MXene so far [15,36]; $Ti_2CT_2$ for comparison with $Ti_3C_2T_2$; and $TiO_2$ as it is the most studied TMO [37]. We consider three different surface functional groups, T= -O/-OH/-F, on the $Ti_2CT_2$ and $Ti_3C_2T_2$ MXenes, because use of LiF+HCl or HF in preparing MXene usually led to some -F functional groups on the surface [38,39]. For $TiO_2$, we focus on the anatase phase since it is more widely used in energy-storage applications than other phases. Below we first elaborate our computational approach.

**II. COMPUTATIONAL DETAILS AND HETEROSTRUCTURE MODELS**

Density functional theory (DFT) calculations were performed using the plane-wave pseudopotential method as implemented in the Vienna *ab initio* simulation package (VASP) [40,41]. The electron-ion interactions were described by the projector augmented-wave (PAW) [42,43] methods while electron exchange-correlation was by the Perdew-Burke-Ernzerhof (PBE) [44] functional form of generalized-gradient approximation (GGA). The kinetic energy cutoff 500 eV was used for the plane-wave basis set. The Grimme DFT-D3 dispersion correction [45] with the Becke-Jonson damping [46,47] was employed to account for the van der Waals (vdW) interactions. We chose the DFT-D3(BJ) method because it has been shown to predict accurate lattice parameters [48] and provide a very good description of MXene surface chemistry [49] as well as graphene-layer stacking [50]. On the other hand, we note that no vdW-included DFT method is truly predictive at this moment [51]; our own test of two ab initio vdW methods (vdW-DF1 and optPBE-vdW) indeed showed variations among the different vdW-DFT methods, which can be found in Table S1 of the supplemental material (SM) [52].

The heterostructure comprises a lateral supercell of the anatase $TiO_2$ (101) surface (the most stable one) matched to a similar lateral supercell of the MXene basal plane. The optimized lattice constants of the tetragonal anatase $TiO_2$ unit cell are **a** = **b** = 3.827 Å and **c** = 9.665 Å, and those for the 2D MXenes are: $Ti_2CO_2$, 3.043 Å; $Ti_2C(OH)_2$, 3.092 Å; $Ti_3C_2O_2$, 3.057 Å; $Ti_3C_2(OH)_2$, 3.105 Å. When constructing the heterostructures, we tried to minimize the lattice mismatch while limiting the size of the supercell. In general, the lattice mismatch becomes smaller as the supercell becomes bigger, but the computational cost also becomes much greater. As a tradeoff, one wants a small-enough lattice mismatch for a not-so-big supercell in building such heterostructures. We found that the constructed heterostructure with surface vectors U=[1,3,-1], V=[0,3,0] for $TiO_2$ and U=[6,3,0], V=[3,4,0] for the MXenes satisfies this tradeoff. **Table I** shows that the constructed heterostructures all have lattice mismatches < 2.5% (following a previous definition [53]) for various surface chemistry of $Ti_2CT_x$ and $Ti_3C_2T_x$. For comparison, we also constructed two different supercells of the heterostructures with using different surface vectors (see Fig. S1 in SM) [52]; both energetics and interlayer spacing (see Table S2 in SM) [52] confirmed that the heterostructure model in Table I is a good representation of the interface.

TABLE I. Lateral lattice parameters and mismatches for heterostructures examined with surface vectors of U=[1,3,-1], V=[0,3,0] for TiO$_2$ and U=[6,3,0], V=[3,4,0] for the MXenes.

| Heterostructure | a [Å] | b [Å] | γ [°] | Lattice mismatch |
|---|---|---|---|---|
| Ti$_2$CO$_2$/TiO$_2$ | 15.651 | 11.227 | 43.029 | 1.8% |
| Ti$_2$CF$_{0.2}$O$_{1.8}$/TiO$_2$ | 15.651 | 11.227 | 43.029 | 1.8% |
| Ti$_2$C(OH)$_2$/TiO$_2$ | 15.777 | 11.314 | 43.029 | 2.0% |
| Ti$_3$C$_2$O$_2$/TiO$_2$ | 15.686 | 11.252 | 43.029 | 1.7% |
| Ti$_3$C$_2$F$_{0.2}$O$_{1.8}$/TiO$_2$ | 15.686 | 11.252 | 43.029 | 1.7% |
| Ti$_3$C$_2$(OH)$_2$/TiO$_2$ | 15.811 | 11.338 | 43.029 | 2.4% |

The slab models for the heterostructures contain three TiO$_2$(101) layers and a MXene layer, with a vacuum layer of 15 Å along the c-axis between the far sides of the heterostructures to avoid spurious interactions. The MXene layer and the two TiO$_2$ layers above it were allowed to relax during the structure optimizations, while the TiO$_2$ layer farthest away from the MXene was kept at their bulk positions. The Brillouin zone was sampled by a converged 5×5×1 Monkhorst-Pack grid [54]; the convergence test can be found in Fig. S2 in SM [52]. More than 400 empty bands were included to make sure that the density of states converges up to 5 eV above the Fermi level; see Fig. S3 in SM for an example of convergence test for the Ti$_3$C$_2$O$_2$/TiO$_2$ heterostructure [52]. Convergence criteria were set to be 0.02 eV/Å for force and 10$^{-5}$ eV for energy. The adhesive energy ($E_{inter}$) of the interface between MXene and TiO$_2$ layers is defined as: $E_{int} = (E_{MXene} + E_{TiO_2} - E_{total})/A$, where $E_{MXene}$, $E_{TiO_2}$, and $E_{total}$ represent the energies of the MXene layer, the TiO$_2$ layer, and the heterostructure, respectively; $A$ is the area of the interface.

### III. RESULTS AND DISCUSSIONS
#### A. Work functions and electronic structure of the heterostructures' building blocks

**Table II** shows the calculated work functions for all the building blocks. One can see that the work functions of all the MXenes are lower than that of the anatase TiO$_2$ (101) surface [a-TiO$_2$ (101)]. Experimental measured work function of a-TiO$_2$ varies with different surface stoichiometries, and the fully oxidized a-TiO$_2$ is around 6.76 eV [55]. As expected, -OH termination leads to the lowest work function [56], while including a fraction of -F groups on the surface lowers the work function of the -O termination. Moreover, there is a small change in the work function from $Ti_2CT_2$ to $Ti_3C_2T_2$. When forming the heterostructures, the vacuum levels align so the building block with a lower work function will transfer electrons to the one with a higher work function. Therefore, one expects that the electrons will flow from MXene to anatase $TiO_2(101)$, no matter what the MXene's termination is but the degree of the charge transfer will depend on the termination group, as will be seen in Sec. III C.

**TABLE II.** Calculated work function (in eV) of the building blocks.

| System | Work Function | System | Work Function |
| --- | --- | --- | --- |
| $Ti_2CO_2$ | 5.70 | $Ti_3C_2O_2$ | 5.94 |
| $Ti_2CF_{0.2}O_{1.8}$ | 5.67 | $Ti_3C_2F_{0.2}O_{1.8}$ | 5.04 |
| $Ti_2C(OH)_2$ | 1.63 | $Ti_3C_2(OH)_2$ | 1.57 |
| anatase-$TiO_2(101)$ | 6.43 | | |

**Fig. 1** compares the DOS of the building blocks. $Ti_2CO_2$ has a small band gap of 0.3 eV, while $Ti_3C_2O_2$ is metallic (Fig. 1a). In addition, by substituting the –F terminated groups for 10% of the -O terminations, the conduction band of $Ti_2CF_{0.2}O_{1.8}$ will move below the fermi level, resulting the metallic character (Fig. 1b). To further analysis the reason that F atoms make the $Ti_2CF_{0.2}O_{1.8}$ metallic, we plot the local density of states (LDOS) for $Ti_2CF_{0.2}O_{1.8}$ in Fig. S4 of SM [52]. We find that the metallic character of $Ti_2CF_{0.2}O_{1.8}$ mainly comes from the occupied conduction band of Ti atoms which is understandable, because F atom could only accept one electron from Ti atom, which is less than O atom. Thus, the Ti atom will have rich electron occupy conduction band, making the structure metallic. In addition, both $Ti_2C(OH)_2$ and $Ti_3C_2(OH)_2$ (Fig. 1c) are metallic; analysis of the orbital contributions indicates that the states at the Fermi level are mainly from Ti 3d. $TiO_2$ is a semiconductor (Fig. 1d): the valence band maximum is of mainly O 2p states and the conduction band minimum Ti 3d. The electronic properties of the building blocks provide a basis for our discussion of the heterostructures as follows. It is well known that the PBE functional significantly underestimates the band gap of bulk $TiO_2$. We think that this is probably less a concern here due to our focus on the interface, the ultrathin nature of the $TiO_2$ layer, and the dominating metallic feature of the MXene layers.

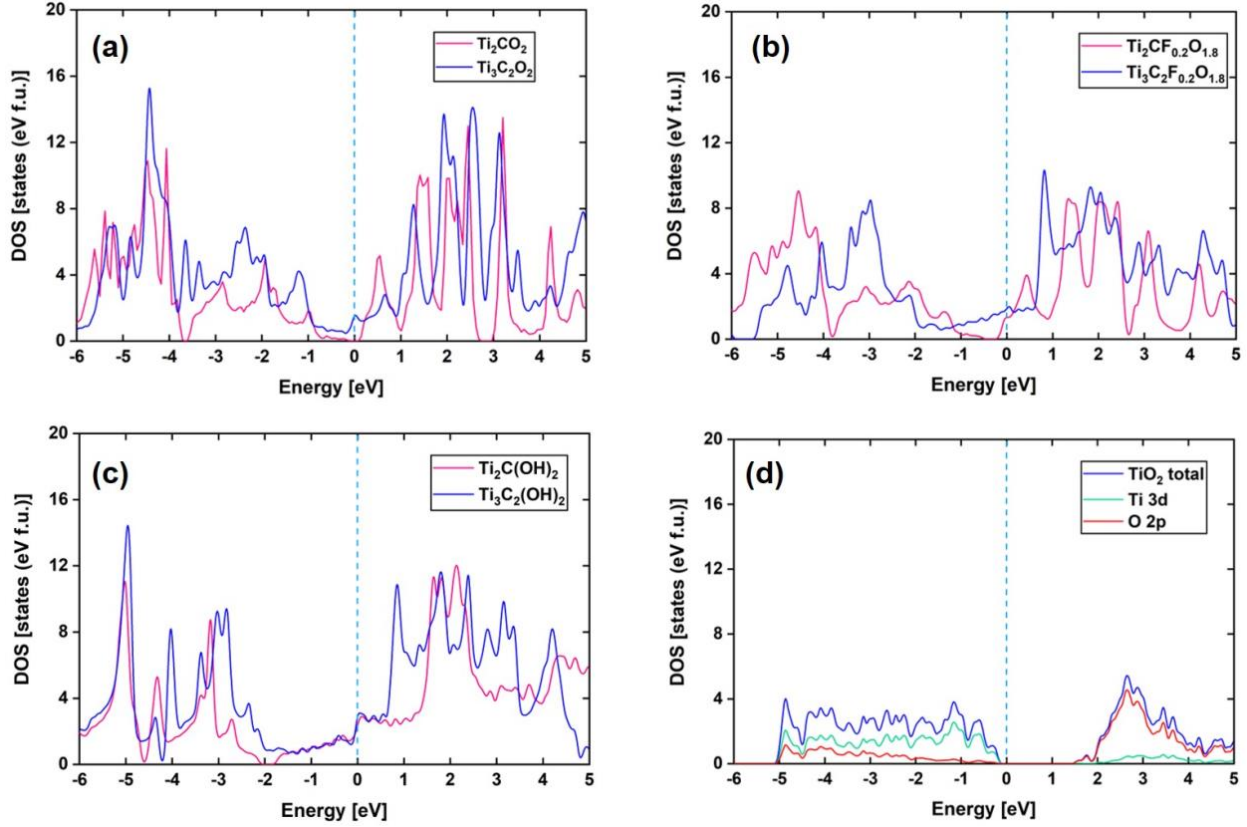

**FIG. 1.** Density of states (DOS) per formula unit (f.u.): (a) $Ti_2CO_2$ and $Ti_3C_2O_2$; (b) $Ti_2CF_{0.2}O_{1.8}$ and $Ti_3C_2F_{0.2}O_{1.8}$; (c) $Ti_2C(OH)_2$ and $Ti_3C_2(OH)_2$; (d) anatase $TiO_2$(101).

### B. Interfacial structure and energetics of the heterostructures

**Fig. 2** shows the optimal configurations for all heterostructures under investigation and their plane-averaged interlayer distances between the surface-O layer from a-$TiO_2$(101) to the O-layer from MXene (namely, the distance between the two closest O-layers at interface). One can see that the MXene thickness ($Ti_2CT_2$ vs. $Ti_3C_2T_2$) or surface chemistry (O vs. OH vs. F) does not impact the plane-averaged interlayer O-O distance much at the interface, which varies from 2.23 Å to 2.28 Å. In the case of -OH termination, this short O-O distance indicates strong hydrogen bonding between -OH of MXene and -O of $TiO_2$. One can see from Fig. 2c,f and insets that multiple H-bonds form at the interface between -OH groups on the MXenes and surface O atoms of $TiO_2$ and that -OH groups and their H atoms on the MXenes rotate and orient towards the surface O atoms on $TiO_2$ to form those H-bonds. This strong interfacial interaction leads to much stronger interfacial adhesion between OH-terminated MXene and $TiO_2$ than between O- or mixed O-/F-termination and TiO$_2$ (**Table III**). After partial F substitution, the interfacial adhesion becomes slightly weaker. Moreover, Ti$_2$CT$_2$ has slightly stronger adhesion with TiO$_2$ than Ti$_3$C$_2$T$_2$ does.

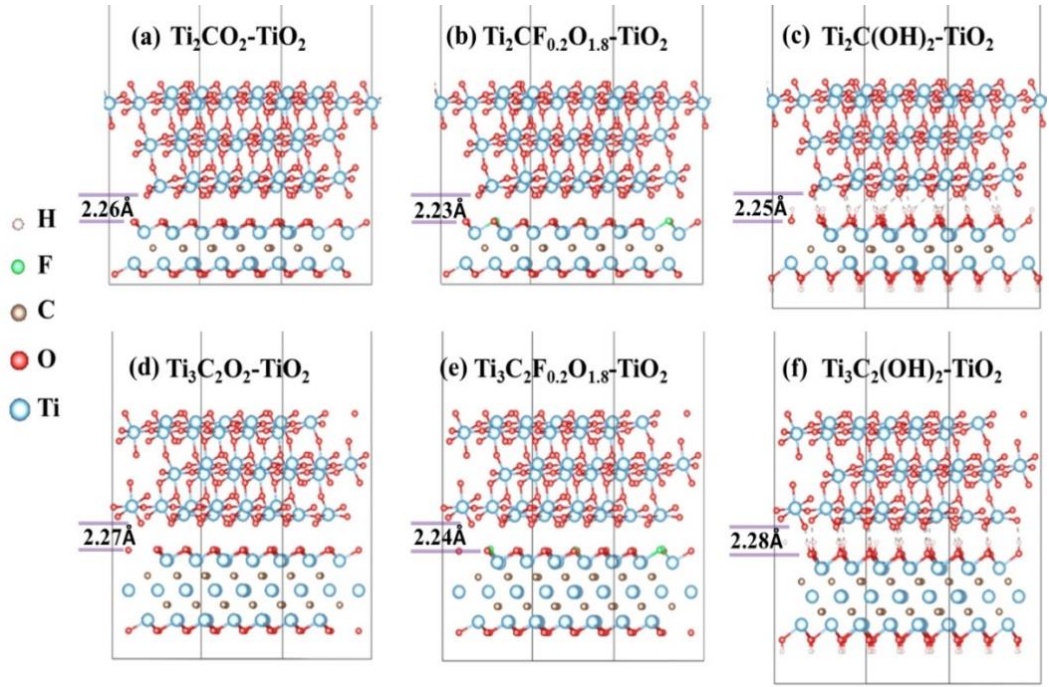

**FIG. 2.** Optimized geometries of different heterostructures. The plane-averaged interlayer distance is measured between the bottom surface-O layer of anatase-TiO$_2$(101) and the top O-layer of MXene.

**TABLE III.** Interfacial adhesion energy ($E_{int}$) for different MXene/TiO$_2$ systems.

| Structure | $E_{int}$ [eV/nm$^2$] |
|---|---|
| Ti$_2$CO$_2$/TiO$_2$ | 1.00 |
| Ti$_2$CF$_{0.2}$O$_{1.8}$/TiO$_2$ | 0.93 |
| Ti$_2$C(OH)$_2$/TiO$_2$ | 5.31 |
| Ti$_3$C$_2$O$_2$/TiO$_2$ | 0.82 |
| Ti$_3$C$_2$F$_{0.2}$O$_{1.8}$/TiO$_2$ | 0.71 |
| Ti$_3$C$_2$(OH)$_2$/TiO$_2$ | 5.11 |

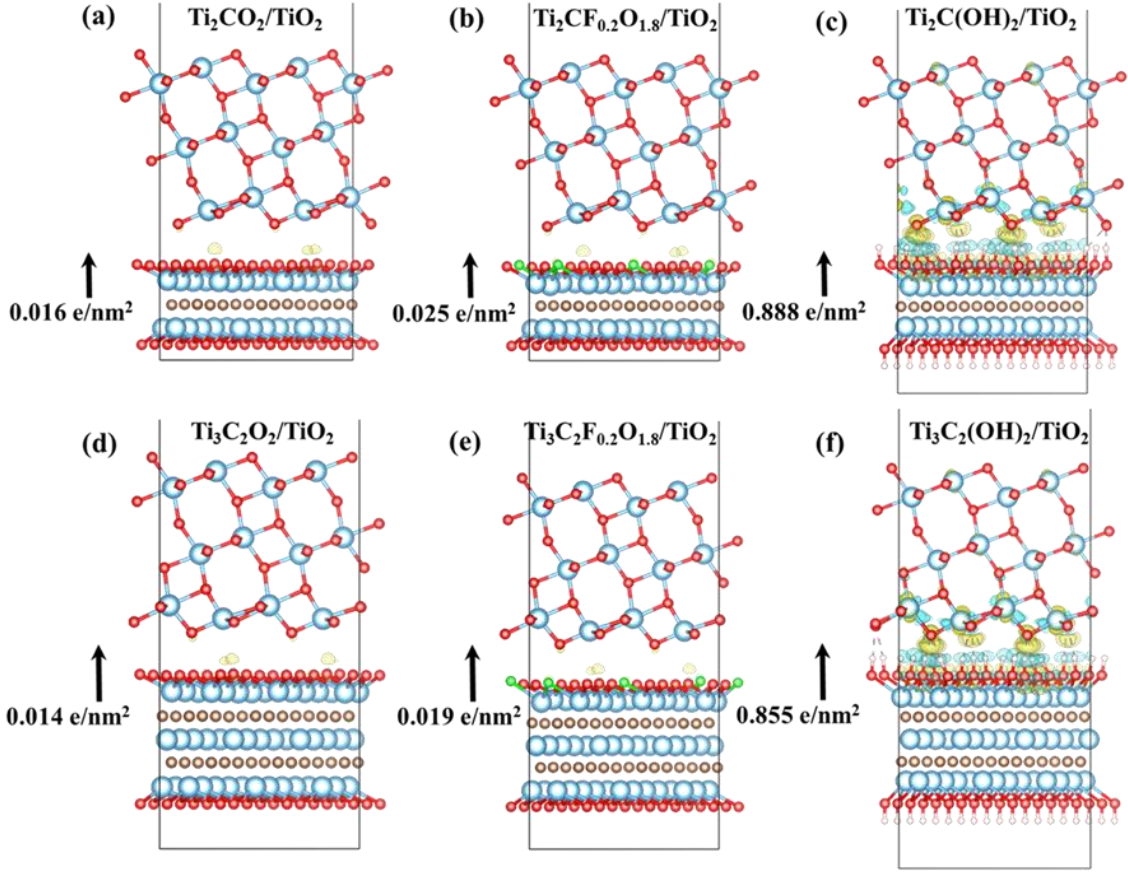

**FIG. 3.** Electron transfer and electron-density-difference isosurfaces for different heterostructures. Arrow and number illustrate the electron transfer direction and amount, respectively. Isosurface value at 0.003 e/Å$^3$: electron accumulation, yellow; electron depletion, cyan.

### C. Interfacial charge transfer of the heterostructures

As explained in Sec. III A, electron transfers from MXene to TiO$_2$ for all terminations, due to the higher work function of TiO$_2$(101) (Table II). **Fig. 3** depicts the electron transfer across the interface for all heterostructures, calculated from the integration of the plane-averaged electron density difference along z-direction. One can see that the amount of electron transfer is much higher for -OH terminations (0.85 – 0.89 e/nm$^2$) than the other terminations (0.014 – 0.025 e/nm$^2$) due to their much lower work functions. The electron-density-difference isosurfaces at the Ti$_2$C(OH)$_2$/TiO$_2$ and Ti$_3$C$_2$(OH)$_2$/TiO$_2$ interfaces clearly show the shift of electron from H on the MXene surface to the O atoms of TiO$_2$ (Fig. 3c,f). To better understand electron distributions within interfaces, the plane-averaged electron-density differences are plotted in **Fig. 4**. One can see that most of the electrons are transferred from the -OH layer of MXene to the surface Ti-O layer of TiO$_2$ at the Ti$_2$C(OH)$_2$/TiO$_2$ and Ti$_3$C$_2$(OH)$_2$/TiO$_2$ interfaces. The electron redistribution is minimal at the heterostructure interfaces of O- or mixed O-/F- termination of MXene.

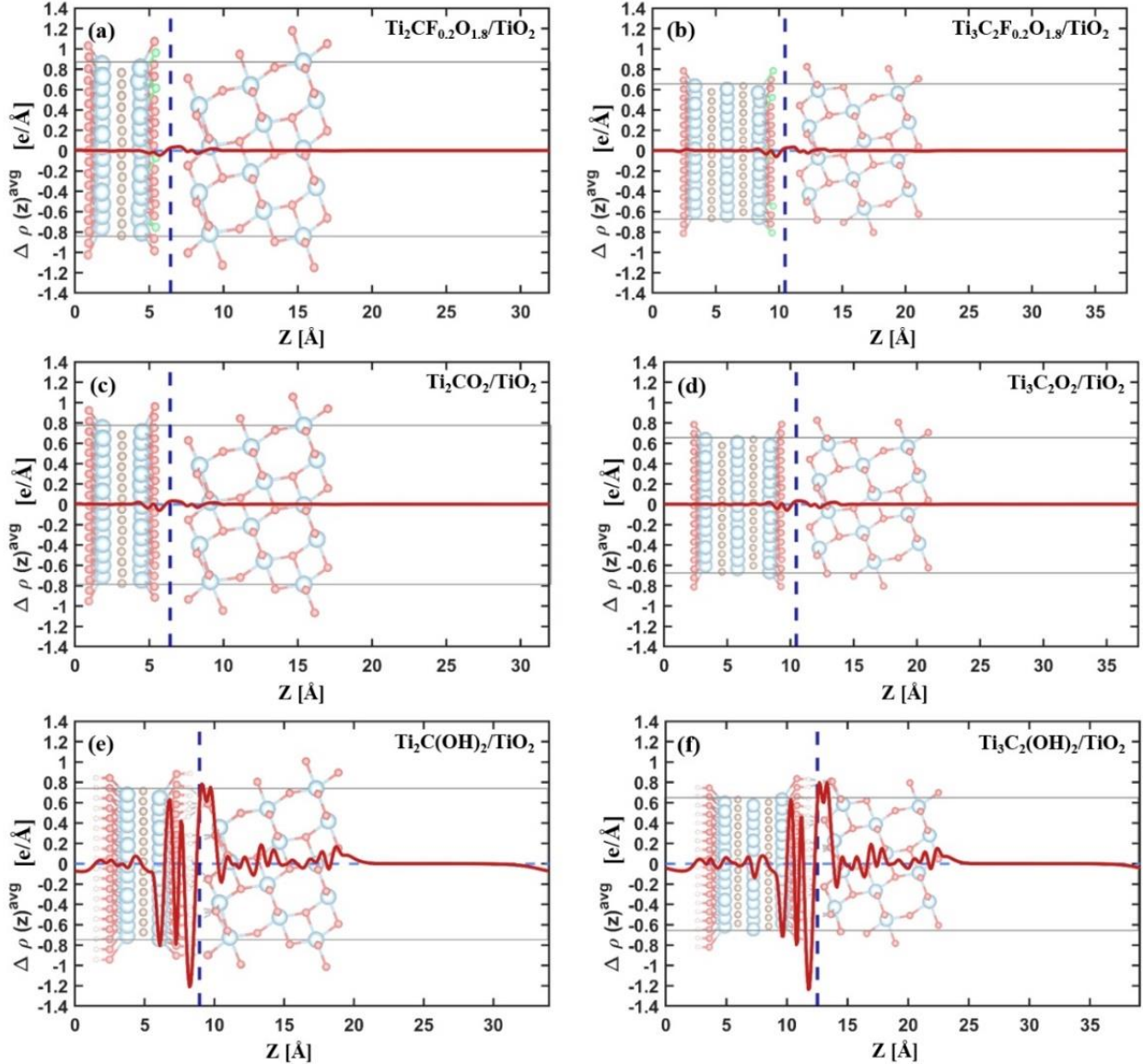

**FIG. 4.** The plane-averaged electron-density difference for different heterostructures. The dashed vertical line represents the plane in the middle of the interface.

### D. Interfacial electronic structure of the heterostructures

We further plot the local density of states (LDOS) for the different heterostructures, projected to their building blocks, in **Fig. 5**. The Ti$_2$CO$_2$/TiO$_2$ heterostructure has a small gap of 0.3 eV (Fig. 5a), similar to that of the pristine Ti$_2$CO$_2$ layer (Fig. 1a). In fact, the LDOS of Ti$_2$CO$_2$ and $TiO_2$ layers in the $Ti_2CO_2/TiO_2$ heterostructure do not change much from those of the isolated building blocks, confirming the minimal electronic interaction between $Ti_2CO_2$ and $TiO_2$ layers in the heterostructure. Similar conclusions can be drawn for $Ti_3C_2O_2/TiO_2$ (Fig. 5b), $Ti_2CF_{0.2}O_{1.8}/TiO_2$ (Fig. 5c), and $Ti_3C_2F_{0.2}O_{1.8}/TiO_2$ (Fig. 5d), which are metallic because the MXene building blocks ($Ti_3C_2O_2$, $Ti_2CF_{0.2}O_{1.8}$, and $Ti_3C_2F_{0.2}O_{1.8}$; Fig. 1a,b). The metallic character of $Ti_2CF_{0.2}O_{1.8}$ stems mainly from the Ti 3d states occupying the conduction band, while the 2p states of F atoms are ~6.5 eV below the Fermi level (see Fig. S4 in SM) [52]. Due to the large charge transfer from MXene to $TiO_2$ in $Ti_2C(OH)_2/TiO_2$ (Fig. 5e) and $Ti_3C_2(OH)_2/TiO_2$ (Fig. 5f) heterostructures, the conduction band of $TiO_2$ get pushed under the Fermi level, resulting in non-zero LDOS at the Fermi level for $TiO_2$ in the heterostructure.

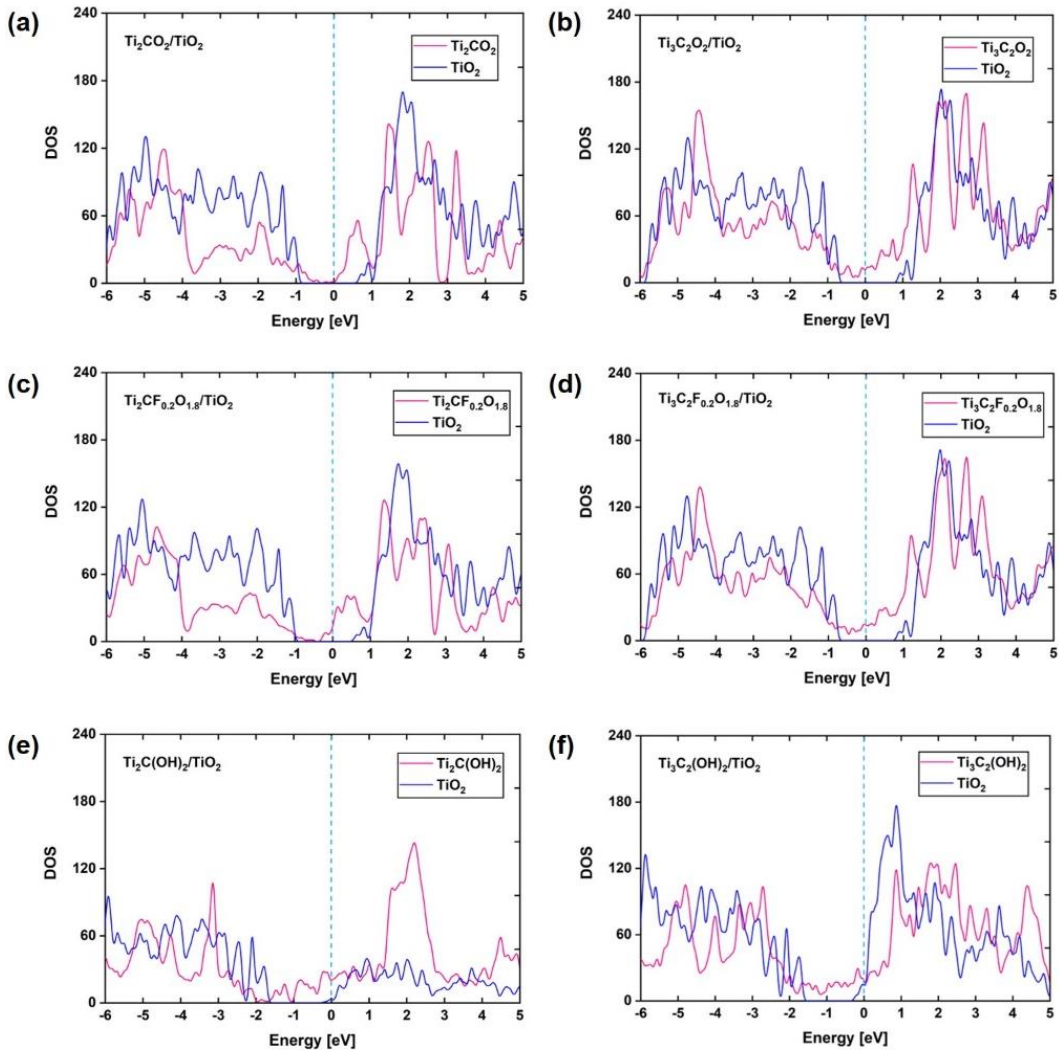

**Fig. 5.** Local density of states (LDOS) of the different heterostructure systems projected to their building blocks: (a) $Ti_2CO_2/TiO_2$; (b) $Ti_3C_2O_2/TiO_2$; (c) $Ti_2CF_{0.2}O_{1.8}/TiO_2$; (d) $Ti_3C_2F_{0.2}O_{1.8}/TiO_2$; (e) $Ti_2C(OH)_2/TiO_2$ and (f) $Ti_3C_2(OH)_2/TiO_2$. The Fermi-level is set to be zero in energy.

Our work here has provided some fundamental insights into a typical MXene/TMO interface. Future efforts include interrogation of ion/electrolyte intercalation into the interface; considerations of different types of MXenes [57] and TMOs; experimental characterization of the interface to confirm the predictions from the present work.

## IV. CONCLUSION

In summary, we have carried out a comprehensive first-principles study of the structural, energetic, and electronic properties of MXene/a-TiO$_2$(101) heterostructures. We found that the greatest work-function differences occur in OH-terminated-MXene/TiO$_2$ interfaces, leading to the largest interfacial electrons transfer (from the OH-terminated MXene to the TiO$_2$ layer) and adhesion. In addition, hydrogen bond formation further strengthens the Ti$_2$C(OH)$_2$/TiO$_2$ and Ti$_3$C$_2$(OH)$_2$/TiO$_2$ interfaces. Electron transfer is minimal and interfacial adhesion is weak for heterostructures of O-, F-, or mixed O-/F- terminations. Our simulations reveal a strong dependence of the interfacial properties of the MXene/TiO$_2$ heterostructures on the surface chemistry of the MXenes which will be useful for tuning the heterostructures for electric-energy-storage needs.

**Acknowledgements**

This research is sponsored by the Fluid Interface Reactions, Structures, and Transport (FIRST) Center, an Energy Frontier Research Center funded by the U.S. Department of Energy (DOE), Office of Science, Office of Basic Energy Sciences. This research used resources of the National Energy Research Scientific Computing Center, a DOE Office of Science User Facility supported by the Office of Science of the U.S. Department of Energy under contract no. DE-AC02-05CH11231.

L.X. and T. W. contributed equally to this work.

**Data Availability**

The data that support the findings of this study are available from the corresponding author upon reasonable request.

## References

[1] E. Pomerantseva and Y. Gogotsi, Nat. Energy **2**, 7 (2017).
[2] C. Choi, D. S. Ashby, D. M. Butts, R. H. DeBlock, Q. Wei, J. Lau, and B. Dunn, Nat. Rev. Mater. **5**, 1 (2020).
[3] J. Li, Z. Du, R. E. Ruther, S. J. An, L. A. David, K. Hays, M. Wood, N. D. Phillip, Y. Sheng, and C. Mao, Jom **69**, 1484 (2017).
[4] M. Sajjad, X. Chen, C. Yu, L. Guan, S. Zhang, Y. Ren, X. Zhou, and Z. Liu, J. Mol. Eng. Mater. **07**, 1950004 (2019).
[5] M. Naguib, V. N. Mochalin, M. W. Barsoum, and Y. Gogotsi, Adv. Mater. **26**, 992 (2014).
[6] J. Come, M. Naguib, P. Rozier, M. W. Barsoum, Y. Gogotsi, P.-L. Taberna, M. Morcrette, and P. Simon, J. Electrochem. Soc. **159**, A1368 (2012).
[7] H. Tang, Q. Hu, M. Zheng, Y. Chi, X. Qin, H. Pang, and Q. Xu, Prog. Nat. Sci. **28**, 133 (2018).
[8] H. Li, Y. Hou, F. Wang, M. R. Lohe, X. Zhuang, L. Niu, and X. Feng, Adv. Energy Mater. **7**, 1601847 (2017).
[9] J. Yan, C. E. Ren, K. Maleski, C. B. Hatter, B. Anasori, P. Urbankowski, A. Sarycheva, and Y. Gogotsi, Adv. Funct. Mater. **27**, 1701264 (2017).
[10] X. Liang, A. Garsuch, and L. F. Nazar, Angew. Chem., Int. Ed. **54**, 3907 (2015).
[11] Q. Tang, Z. Zhou, and P. Shen, J. Am. Chem. Soc. **134**, 16909 (2012).
[12] M. Khazaei, A. Ranjbar, M. Arai, T. Sasaki, and S. Yunoki, J. Mater. Chem. C **5**, 2488 (2017).
[13] N. Kurra, B. Ahmed, Y. Gogotsi, and H. N. Alshareef, Adv. Energy Mater. **6**, 1601372 (2016).
[14] E. Pomerantseva, F. Bonaccorso, X. Feng, Y. Cui, and Y. Gogotsi, Science **366**, (2019).
[15] B. Anasori, M. R. Lukatskaya, and Y. Gogotsi, Nat. Rev. Mater. **2**, 1 (2017).
[16] M.-Q. Zhao, C. E. Ren, Z. Ling, M. R. Lukatskaya, C. Zhang, K. L. Van Aken, M. W. Barsoum, and Y. Gogotsi, Adv. Mater. **27**, 339 (2015).
[17] Z. Ling, C. E. Ren, M.-Q. Zhao, J. Yang, J. M. Giammarco, J. Qiu, M. W. Barsoum, and Y. Gogotsi, Proc. Natl. Acad. Sci. **111**, 16676 (2014).
[18] M. Naguib, M. Kurtoglu, V. Presser, J. Lu, J. Niu, M. Heon, L. Hultman, Y. Gogotsi, and M. W. Barsoum, Adv. Mater. **23**, 4248 (2011).
[19] N. C. Osti, M. Naguib, K. Ganeshan, Y. K. Shin, A. Ostadhossein, A. C. Van Duin, Y. Cheng, L. L. Daemen, Y. Gogotsi, E. Mamontov, and A. I. Kolesnikov, Phys. Rev. Mater. **1**, 065406 (2017).
[20] Y.-T. Liu, P. Zhang, N. Sun, B. Anasori, Q.-Z. Zhu, H. Liu, Y. Gogotsi, and B. Xu, Adv. Mater. **30**, 1707334 (2018).
[21] C. Zhang, M. Beidaghi, M. Naguib, M. R. Lukatskaya, M.-Q. Zhao, B. Dyatkin, K. M. Cook, S. J. Kim, B. Eng, and X. Xiao, Chem. Mater. **28**, 3937 (2016).
[22] R. B. Rakhi, B. Ahmed, D. Anjum, and H. N. Alshareef, ACS Appl. Mater. Interfaces **8**, 18806 (2016).
[23] B. Ahmed, D. H. Anjum, M. N. Hedhili, Y. Gogotsi, and H. N. Alshareef, Nanoscale **8**, 7580 (2016).
[24] Y. Aierken, C. Sevik, O. Gülseren, F. M. Peeters, and D. Çakır, J. Mater. Chem. A **6**, 2337 (2018).
[25] Y.-T. Du, X. Kan, F. Yang, L.-Y. Gan, and U. Schwingenschlögl, ACS Appl. Mater. Interfaces **10**, 32867 (2018).
[26] R. Li, W. Sun, C. Zhan, P. R. Kent, and D. Jiang, Phys. Rev. B **99**, 085429 (2019).
[27] I. Demiroglu, F. M. Peeters, O. Gülseren, D. Çakır, and C. Sevik, J. Phys. Chem. Lett. **10**, 727 (2019).
[28] P. Paul, P. Chakraborty, T. Das, D. Nafday, and T. Saha-Dasgupta, Phys. Rev. B **96**, 035435 (2017).
[29] P. Chakraborty, T. Das, D. Nafday, L. Boeri, and T. Saha-Dasgupta, Phys. Rev. B **95**, 184106 (2017).
[30] P. Chakraborty, T. Das, and T. Saha-Dasgupta, Compr. Nanosci. Nanotechnol. **1**, 319. (2019).
[31] J. Huang, R. Meng, L. Zu, Z. Wang, N. Feng, Z. Yang, Y. Yu, and J. Yang, Nano Energy **46**, 20 (2018).
[32] B. Ahmed, D. H. Anjum, Y. Gogotsi, and H. N. Alshareef, Nano Energy **34**, 249 (2017).
[33] M.-Q. Zhao, M. Torelli, C. E. Ren, M. Ghidiu, Z. Ling, B. Anasori, M. W. Barsoum, and Y. Gogotsi, Nano Energy **30**, 603 (2016).
[34] Mashtalir, M. Naguib, V. N. Mochalin, Y. Dall'Agnese, M. Heon, M. W. Barsoum, and Y. Gogotsi, Nat. Comm. **4**, 1 (2013).

[35] Y. Xie and P. R. C. Kent, Phys. Rev. B **87**, 235441 (2013).
[36] M. Alhabeb, K. Maleski, B. Anasori, P. Lelyukh, L. Clark, S. Sin, and Y. Gogotsi, Chem. Mater. **29**, 7633 (2017).
[37] A. J. Haider, Z. N. Jameel, and I. H. M. Al-Hussaini, Energy Procedia **157**, 17 (2019).
[38] M. A. Hope, A. C. Forse, K. J. Griffith, M. R. Lukatskaya, M. Ghidiu, Y. Gogotsi, and C. P. Grey, Phys. Chem. Chem. Phys. **18**, 5099 (2016).
[39] T. Kobayashi, Y. Sun, K. E. Prenger, D.-E. Jiang, M. Naguib, and M. Pruski, J. Phys. Chem. C (2020).
[40] G. Kresse and J. Furthmüller, Phys. Rev. B **54**, 11169 (1996).
[41] G. Kresse and J. Furthmüller, Comput. Mater. Sci. **6**, 15 (1996).
[42] P. E. Blöchl, Phys. Rev. B **50**, 17953 (1994).
[43] G. Kresse and D. Joubert, Phys. Rev. B **59**, 1758 (1999).
[44] J. P. Perdew, K. Burke, and M. Ernzerhof, Phys. Rev. Lett. **77**, 3865 (1996).
[45] S. Grimme, S. Ehrlich, and L. Goerigk, J. Comput. Chem. **32**, 1456 (2011).
[46] A. D. Becke and E. R. Johnson, J. Chem. Phys. **123**, 154101 (2005).
[47] S. Grimme, J. Antony, S. Ehrlich, and H. Krieg, J. Chem. Phys. **132**, 154104 (2010).
[48] M. Fischer, W. J. Kim, M. Badawi and S. Lebègue, J. Chem. Phys. **150**, 094102 (2019).
[49] X. Wang, C. Wang, S. Ci, Y. Ma, T. Liu, L. Gao, P. Qian, C. Ji and Y. Su, J. Mater. Chem. A **8**, 23488 (2020).
[50] I. V. Lebedeva, A.V. Lebedev, A. M. Popov, and A. A. Knizhnik, Comput. Mater. Sci. **128**, 45 (2017).
[51] F. Schulz, P. Liljeroth, and A. P. Seitsonen, Phys. Rev. Mater. **3**, 084001 (2019).
[52] See Supplemental Material at http://xxx. for calculated interlayer distance and interfacial adhesion energy of $Ti_2CO_2/TiO_2$ supercells with different vdW methods, constructed $Ti_2CO_2/TiO_2$ supercells with different surface vectors and its corresponding lattice information, k-grid convergence and empty band tests of $Ti_3C_2O_2/TiO_2$ heterostructure, and the local density of states of $Ti_2CF_{0.2}O_{1.8}$.
[53] A. Christensen and E.A. Carter, J. Chem. Phys. **114**, 5816 (2001).
[54] H. J. Monkhorst and J. D. Pack, Phys. Rev. B **13**, 5188 (1976).
[55] S. Kashiwaya, J. Morasch, V. Streibel, T. Toupance, W. Jaegermann, and A. Klein, Surfaces **1**, 73 (2018).
[56] M. Khazaei, M. Arai, T. Sasaki, A. Ranjbar, Y. Liang, and S. Yunoki, Phys. Rev. B **92**, 075411 (2015).
[57] J. D. Gouveia, F. Viñes, F. Illas, and J. R. Gomes, Phys. Rev. Mater. **4**, 054003 (2020).